\documentclass{aa}
\usepackage{graphicx}
\begin{document}
   \title{Formation of massive stars by growing accretion rate}
   \author{R.~Behrend
          \and
          A.~Maeder
          }

   \institute{Geneva Observatory,
              CH-1290 Sauverny, Switzerland\\
              email: raoul.behrend@obs.unige.ch\\
	      email: andre.maeder@obs.unige.ch}

   \date{Received 3 April 2001 / Accepted 12 April 2001}

   \abstract{
      We perform calculations of pre-main sequence evolution 
      of stars from $1$ to $85\,M_{\odot}$ with growing accretion rates
      $\dot{M}$. The values of $\dot{M}$
      are taken equal to a constant fraction $\tilde{f}$ of the rates of
      the mass outflows observed by Churchwell (\cite{church}) and Henning
      (\cite{henning2000}). The evolution of the various stellar parameters is 
      given, as well as the evolution of the disc luminosity;
      electronic tables are provided as a supplement to the articles. Typically, the
      duration of the accretion phase of massive stars is 
      $\simeq 3 \cdot 10^5\,\mathrm{yr}$. and there is less than $10\%$
      difference in the time necessary to form a $8$ or $80\,M_{\odot}$ star.
      If in a young cluster all the proto--stellar cores start to 
      accrete at the same time, we then have a relation $M(t)$
      between the masses of the new stars and the time $t$ of their 
      appearance. Since we also know the distribution of stellar masses at the end of star
      formation (IMF), we can derive the star formation history
      $N(t)$. Interestingly enough, the current IMF implies two
      peaks of star formation: low mass stars form first and high mass
      star form later.
     }
   \authorrunning{R.~Behrend \and A.~Maeder}
   \titlerunning{Star formation by growing accretion rate}
   \maketitle


\section{Introduction}

Two scenarios have been proposed to explain the formation of massive stars.
Star coalescence (Bonnell et al. \cite{bonnell}; Stahler et al. \cite{stahler})
and the accretion scenario (Beech \& Mitalas \cite{beemit}; Bernasconi \& Maeder \cite{bermae}).
The merging of stars requires very high stellar densities
($\ge 10^4\,\mathrm{stars}\,{\mathrm{pc}^{-3}}$) to be efficient (Bonnell et al.
\cite{bonnell}; Henning \cite{henning}).

The imaging of new-born stars with the HST (Burrows et al. \cite{burrows}), the IRAM
interferometer (Dutrey \cite{dutrey}), the VLA (Wilner \& Ho \& Kastner \& Rodr\'{i}guez
\cite{wilneretal}) and the VLT (Brandner \cite{brandner}), among others)
clearly shows discs (or remnants of discs), favouring the accretion scenario for low and
intermediate mass stars. The jets and/or outflows that are sometime observed are closely related
to accretion discs; they can contribute to solve the angular momentum problem (Tomisaka \cite{tomisaka2}).

Accretion rates of the order of $10^{-5}$ to $10^{-4}\,M_{\sun}\,\mathrm{yr}^{-1}$
explain very well low mass stars (Palla \cite{palla}). But such values fail to describe the formation
of massive stars, because with such rates the massive stars would leave the ZAMS 
(i.e. would have burned a significant part of their hydrogen) before being fully formed,
as pointed out by Nakano (\cite{nakaal3}) and Norberg \& Maeder (\cite{normae}).
Nakano (\cite{nakaal2}) studied the implications of a huge constant accretion rate of $\approx 10^{-2}\,M_{\sun}\,\mathrm{yr}^{-1}$
for an observed $10^5\,L_{\sun}$ protostar.

The reality could well lie between these two models. Can an accretion rate that depends
on the stellar mass (or luminosity) describe the star formation from low to very high masses ?
This is the question of the growing accretion rates.

A nice observational relation between the outflow mass rates and the
stellar bolometric luminosities in ultra compact $\ion{H}{ii}$ regions was established by
Churchwell
(\cite{church}) over the ranges $1$ to $10^6\,L_{\sun}$ and $10^{-6}$ to
$10^{-2}\,M_{\sun}\,\mathrm{yr}^{-1}$. Independent observations by Henning et al.
(\cite{henning2000}) confirm its validity.

Churchwell's relation can bring some new constraints for the scenarios of the formation of massive
stars. A crucial question is to know if the accretion scenario is compatible or not with this relation.
The relation can bring something new to the description of the pre-main sequence evolution of stars
and for the explanation of the rapid birth of massive stars.
The relation can also be a used as a strong check for the computer simulations of discs
around stars and the jets/outflows, and offers the possibility to fix uncertain parameters
of these models (e.g. viscosity).

The principal properties of stars with accretion discs on the birthline are discussed.
Tracks for stars leaving the birthline are calculated.
Links between the IMF and the processes that end the stellar mass evolution are also
presented.


\section{The simulations}

\begin{figure*}
   \includegraphics[width=17cm]{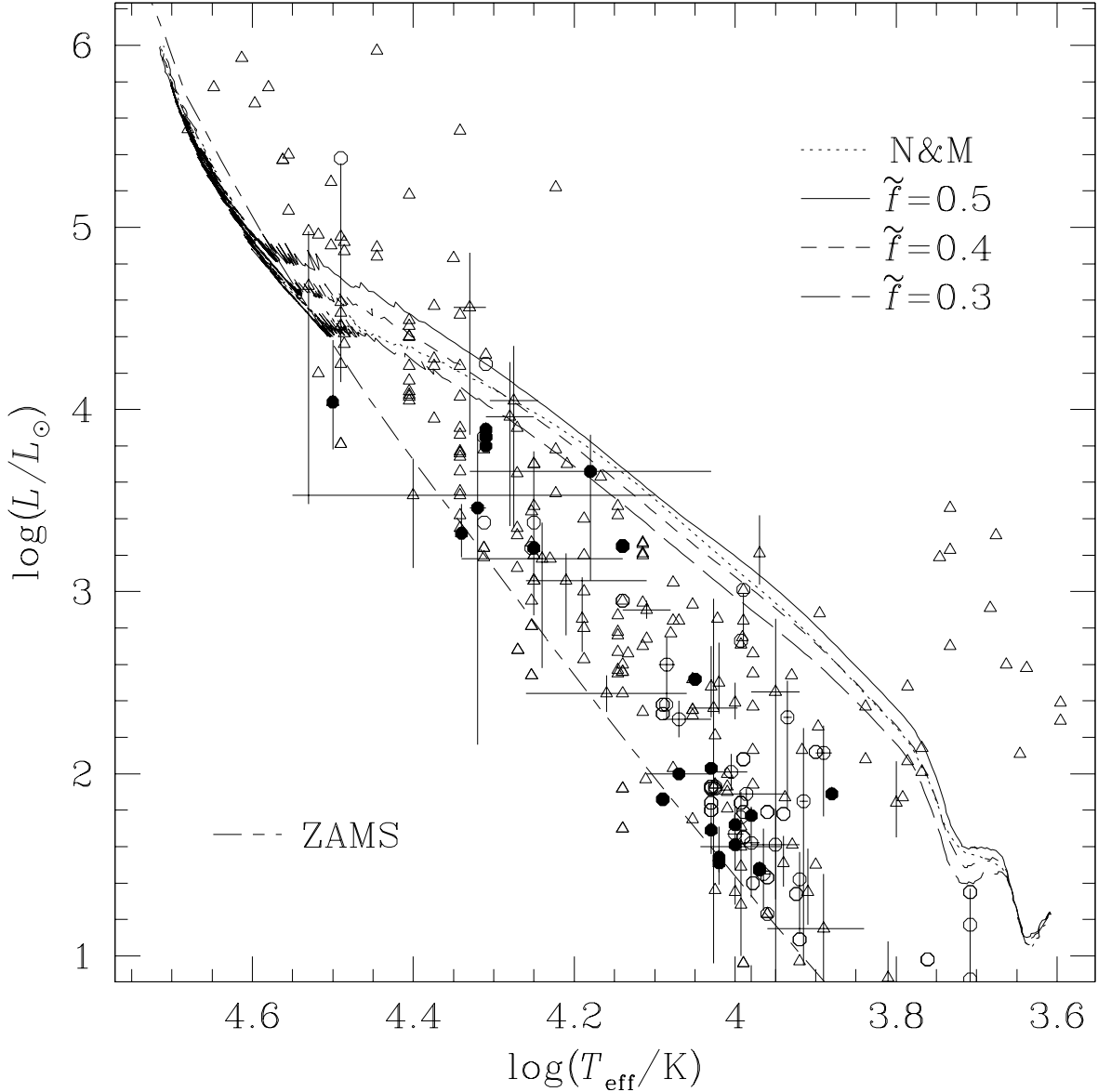}
   \caption{The $\log T_\mathrm{eff}$ -- $\log L$ diagram showing birthlines for 3 values of
   $\tilde{f}$ and for the model of N\&M. The ZAMS is taken
   from Schaller et al. (\cite{schall}). Observations of protostars from the compilation
   by N\&M are shown with discs (respectively circles) if $\dot{M}$ is more
   (respectively less) than $10^{-5}\,M_{\sun}\,\mathrm{yr}^{-1}$; triangles mean unspecified rates.}
   \label{1348f1}
\end{figure*}

The matter which escapes from the disc (in the central region) is divided in two major flows:
one part falls on the star while the other part is ejected in the form of jets and outflows.
Let us denote by $\dot{M}_\mathrm{disc}$ the mass rate transferred inwards from the disc, by $\dot{M}$
the mass rate accreted by the star, and by $\dot{M}_\mathrm{out}$ the mass rate in the jets/outflow.

The relation, found by Churchwell (\cite{church}) and confirmed by Henning et al.
(\cite{henning2000}), between the bolometric luminosity $L$ of
stars and the outflow mass rate $\dot{M}_{\mathrm{out}}$ is single valued
(at least in the error bars), monotonously increasing and valid over a very wide
range of luminosity: from $1$ to $10^6\, L_{\sun}$.
A polynomial fit of the observations by Churchwell (\cite{church}) gives
\[\log{{\dot{M}_\mathrm{out}} \over {M_{\sun}\,\mathrm{yr}^{-1}}}=-5.28+\log {L
\over{L_{\sun}}} \cdot(0.752-0.0278\log {L \over{L_{\sun}}}).\]

Unfortunately, there is at present no similar relation available between $L$ and $\dot{M}$;
one must parameterize the ratio between the accretion and outflow fluxes and do some suppositions
about their proportion before being able to model the evolution of the central star.
Let be $\dot{M}=\tilde{f}\,\dot{M}_\mathrm{out}$, $\dot{M}=f\,\dot{M}_\mathrm{disc}$ and
$\dot{M}_\mathrm{out}=(1-f)\,\dot{M}_\mathrm{disc}$ so that $\tilde{f}=f/(1-f)$.
A variety of physical parameters could have effects on $\tilde{f}$ (or $f$): chemical compositions, metallicities,
magnetic fields, rotational velocity of the star. Some other parameters have
certainly a role to play, but are not yet firmly defined, like the dust opacity.
Therefore, obtaining a theoretical value of $f$ is very difficult. Nonetheless, Tomisaka (\cite{tomisaka1})
found $f \approx 1/3$ with his numerical MHD-simulation of a low mass class $0$ protostar.
Shu et al. (\cite{shuetal}) deduced a similar value with their model of X-wind magnetic
configuration, also for a low mass star. From his measurements of the far-infrared
luminosity of a B star and the mass of its outflows, Churchwell (\cite{church}) estimated
a slightly smaller value, of the order of $f \approx 0.15$.

Due to the great number of changing physical parameters which can have an effect on $f$, $f$ is
probably non-constant during the evolution of accreting stars. The last three values of
$f$ cited above suggest that $f$ could be a decreasing function of the stellar mass, but this is
far from being firmly established. For this preliminary study, we simplify
this huge astrophysical problem, retaining only a ratio $f$ that is constant during the whole
accretion period.

The "best" $f$ is presently unknown. A way to assign a more or less realistic value to $f$ is as follows.
The birthline is the path in the HR diagram that continuously accreting stars follow.
Stars on the birthline are difficult to observe,
because they are hidden by the accretion disc and a dense cocoon of interstellar matter. The star becomes
visible once a sufficient amount of the surrounding matter has been dispersed.
It follows that the young stars are observed generally between the birthline and the zero
age main-sequence (ZAMS). Before hydrogen ignition, gravitational
contraction and deuterium burning are the main energy sources for the star; the
localization of the birthline depends on the deuterium accretion rate. We will choose $\tilde{f}$
as the value for which the
computed birthline fits "at the best" the upper envelope of the observations of young stars in
the pre-main sequence part of the HR diagram (as in Norberg \& Maeder \cite{normae}, denoted hereafter N\&M).

The evolution of the pre-main sequence models with growing accretion rate 
is computed with the Geneva code (Meynet \& Maeder \cite{meymae}), under
the assumption of a non-rotating star, at $Z=0.02$ and at deuterium concentration
$5 \cdot 10^{-5}$. Winds are not taken into account.

The computations begin with a fully convective $0.7\,M_{\sun}$ protostar which is taken
to be $7 \cdot 10^5\,\mathrm{yr}$ old, in agreement with models of low mass star
formation (Palla \cite{palla}). The mass of the protostar grows at a rate
$\dot{M}=\max(10^{-5}\,M_{\sun}\,\mathrm{yr}^{-1},\,\tilde{f}\,\dot{M}_\mathrm{out})$.

We compare the models with accretion rates parameterized by the bolometric luminosity of
the star to the model of mass parameterized accretion rate found by N\&M:
$\dot{M}=10^{-5}\cdot\max(1,\,M)^{1.5}$ where units are $M_{\sun}$ and
$M_{\sun}\,\mathrm{yr}^{-1}$.
The $M$- and $L$-parameterized accretion rates are different in the sense that the
first one depends only on one parameter (the mass) of the star, and the second one depends
on many involved quantities (mass, profiles of chemical composition and internal energy,
etc.) and thus is sensitive to the history of the star (e.g. opacity via chemical abundances,
diffusive processes). For this reason, the memory of the initial luminosity of the star in
the $M$-parameterization is rapidly lost. This is not the case for the $L$-parameterized
accretion; if the initial star is over-/under-luminous, the luminosity of the accreted
deuterium is also above/below the normal value and the birthline is also
over-/under-luminous. This sensitivity to the initial conditions for the $L$-parameterized
accretion remains small with respect to the possible variations of the birthlines with $f$,
for stars initially on the Hayashi band.
Processes that depend on their past can "easily" oscillate or even be chaotic, for physical
reasons (Hairer et al. \cite{haireral}) or for numerical reasons (Berg\'{e} et al. \cite{bergal}).
Apparently the calculations do not reveal strong signatures of instabilities, except maybe
some oscillations near the surface of accreting stars above $22\,M_{\sun}$.


\section{Results and discussions}

\begin{figure}
   \includegraphics[width=8.8cm]{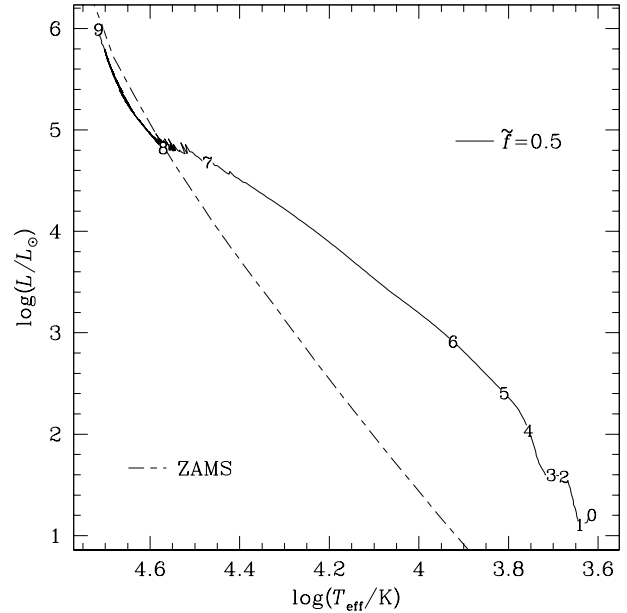}
   \caption{A birthline in $\log T_\mathrm{eff}$ -- $\log L$. Some key
   phases are highlighted:
   $0$: initial conditions
   - $1$: central ignition of deuterium
   - $2$: central gravitation dominates over $\ion{^2H}{}$-burning
   - $3$: start of shell deuterium-burning
   - $4$: end of large scale convection
   - $5$: end of thin central convection
   - $6$: central $\ion{^2H}{}$-exhaustion
   - $7$: central ignition of hydrogen and restart of core convection
   - $8$: arrival on the ZAMS
   - $9$: star starts evolving away from the ZAMS.}
   \label{1348f2}
\end{figure}

The first step consists of the estimate of a plausible $\tilde{f}$.
The second step is the prediction of physical parameters for the stars and discs using
the $\tilde{f}$ found. Connections with the IMF is postponed to the next section.

Before hydrogen ignition, the main energy sources for the star are gravitational
contraction and deuterium burning. The localization of the birthline is then governed
essentially by the deuterium accretion rate on the star. The higher $\dot{M}$, the higher the
birthline in the HR diagram.

\begin{table}[htb]
   \caption[]{Properties of stars on the birthline with $\tilde{f}=0.5$:
   age, luminosity, mass, effective temperature, radius, relative upper radius of the
   convective zone. The ratio of disc and star luminosities is also given.}
   \begin{center}
      $\begin{array}{ccccccc}
          t/\mathrm{yr}     & L/L_{\sun} & M/M_{\sun} & T_\mathrm{eff}/\mathrm{K} & R/R_{\sun} &
          R_\mathrm{conv}^\mathrm{sup} / R & L_\mathrm{disc}/L \\[1mm]
           70.01e3 &  16.00     &  0.700 &  4110 &  7.88 & 1.000 &  4.17 \\
           74.50e3 &  15.14     &  0.785 &  4176 &  7.43 & 1.000 &  6.22 \\
           81.20e3 &  13.24     &  0.901 &  4290 &  6.58 & 1.000 &  8.33 \\
           88.45e3 &  12.82     &  1.02  &  4377 &  6.22 & 1.000 &  9.80 \\
           98.13e3 &  15.41     &  1.19  &  4442 &  6.63 & 1.000 &  9.75 \\
          114.5e3  &  23.30     &  1.55  &  4533 &  7.82 & 1.000 &  9.47 \\
          130.1e3  &  33.25     &  1.99  &  4639 &  8.92 & 1.000 &  9.67 \\
          159.0e3  &  38.91     &  2.98  &  4890 &  8.68 & 1.000 & 14.6  \\
          188.2e3  &  38.91     &  3.99  &  5112 &  7.95 & 1.000 & 21.4  \\
          217.0e3  &  48.80     &  5.05  &  5358 &  8.10 & 1.000 & 24.0  \\
          232.8e3  & 379.1      &  6.01  &  7036 & 13.1  & 0.000 &  8.32 \\
          239.8e3  &   2.662e3  &  8.05  & 11.73e3  & 12.5  & 0.000 &  5.21 \\
          242.8e3  &   9.100e3  & 10.0   & 16.55e3  & 11.6  & 0.000 &  4.00 \\
          244.6e3  &  21.58e3   & 12.0   & 21.79e3  & 10.3  & 0.000 &  3.55 \\
          246.2e3  &  46.77e3   & 15.0   & 29.76e3  &  8.13 & 0.001 &  3.92 \\
          248.3e3  &  81.19e3   & 20.1   & 37.74e3  &  6.66 & 0.135 &  5.17 \\
          250.2e3  &  87.63e3   & 25.0   & 39.40e3  &  6.35 & 0.229 &  6.65 \\
          251.9e3  & 132.1e3    & 30.0   & 42.61e3  &  6.66 & 0.234 &  5.50 \\
          254.5e3  & 256.7e3    & 40.0   & 46.59e3  &  7.77 & 0.265 &  4.85 \\
          256.4e3  & 372.3e3    & 49.9   & 48.17e3  &  8.75 & 0.272 &  3.82 \\
          258.1e3  & 495.1e3    & 59.7   & 49.37e3  &  9.61 & 0.288 &  3.73 \\
          259.7e3  & 680.4e3    & 70.6   & 50.70e3  & 10.7  & 0.299 &  3.31 \\
          260.9e3  & 825.5e3    & 79.4   & 51.49e3  & 11.4  & 0.305 &  3.02 \\
          261.6e3  & 952.7e3    & 84.9   & 51.84e3  & 12.0  & 0.312 &  2.87

      \end{array}$							  
   \end{center}
   \label{tabbirthline}
\end{table}

\begin{figure*}
   \includegraphics[width=17cm]{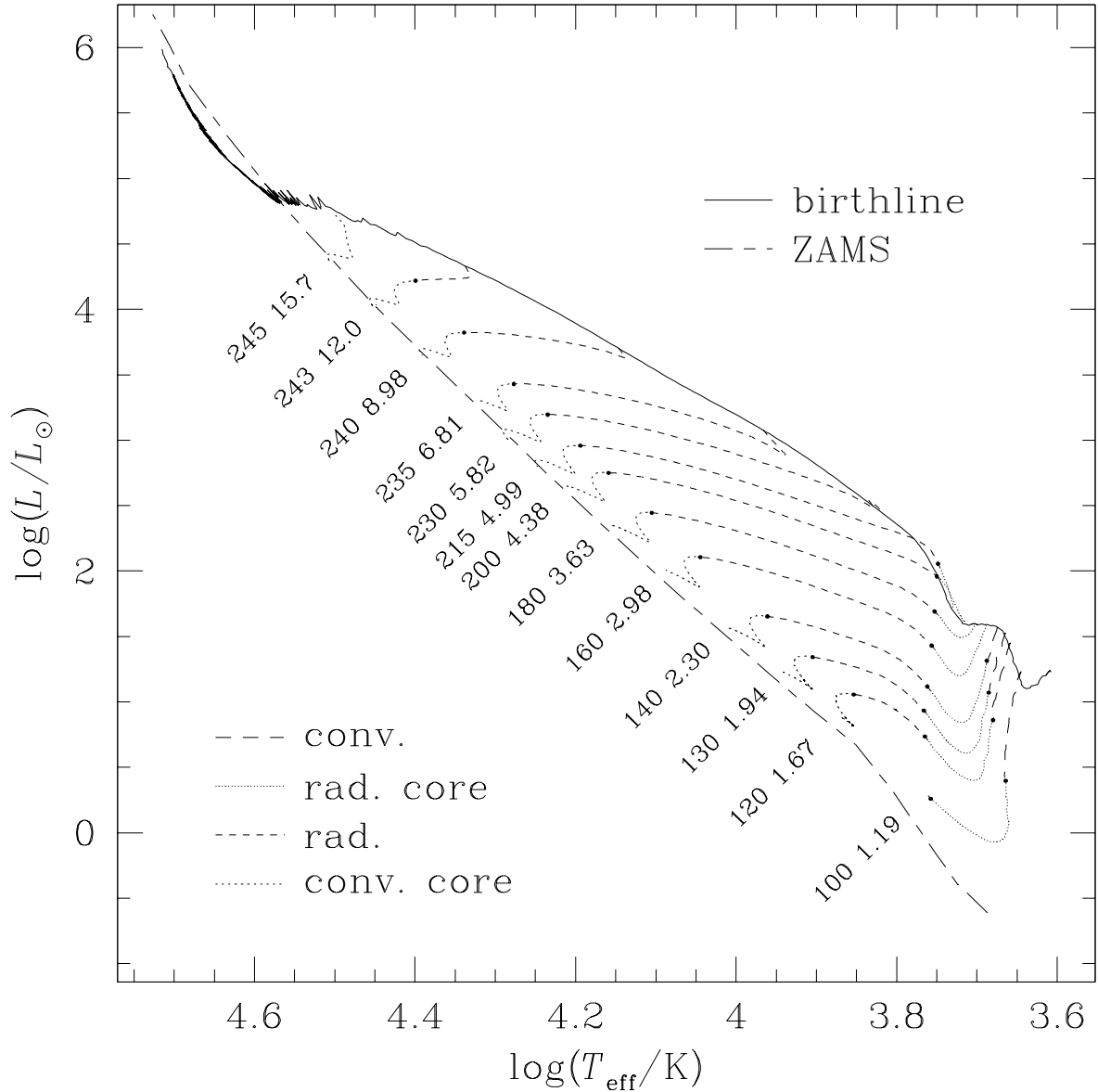}
   \caption{Tracks in $\log T_\mathrm{eff}$ -- $\log L$ followed by stars that leave the
   birthline with $\tilde{f}=0.5$. Couples of numbers represent
   the age $t_\mathrm{stop}$ of the star (in units of $10^3\,\mathrm{yr}$) at the stop of
   the accretion, and its mass (in $M_{\sun}$) on the ZAMS. For the stars which are no
   longer on
   the birthline, the small dots at $\log T_{\mathrm{eff}} \approx 3.75$ correspond to the
   beginning of the fully radiative interiors; the small dots close to the ZAMS correspond
   to the appearance of the convective cores.}
   \label{1348f8}
\end{figure*}

Figure~\ref{1348f1} shows birthlines for different values of $\tilde{f}$, and the
observations compiled in N\&M.
While $\tilde{f}=0.3$ seems to be a little too low, $\tilde{f} \in \{0.4,\,0.5\}$ give reasonable
envelopes to the observations.
Two statistical biases in the estimate of the envelope are possible, with consequences
for the determination of $\tilde{f}$.
1) The stars which constrain the birthline are usually all on one side of the birthline (a
nonsignificant exception is nevertheless visible in Fig.~\ref{1348f8}).
Because the density of stars near the birthline is small, the probability that one misses a
crucial observation is relatively large. The effect of this bias is an underestimated
luminosity of the envelope of the observation and thus also of the birthline. Unfortunately, this occurs in
the upper pre-main sequence part of the HR diagram where the sensibility of the computed
birthline to $\tilde{f}$ is the greatest. The deficiency of stars in this region is due to the
very short time spent by a star in this region, as will be shown in Fig.~\ref{1348f8}.
2) The stars leave the birthline faster than they reach the ZAMS. In other words, the median of the
positions of the stars for which the error bars constrain the birthline (by their envelope)
is shifted towards the ZAMS; this effect vanishes as error bars go to zero.
With these two facts in mind, our preference is for the highest value: $\tilde{f}=0.5$,
i.e. $f=1/3$.

The ratio $f=\dot{M}_{\mathrm{out}}/\dot{M}_{\mathrm{disc}}$ supported by Fig.~\ref{1348f1} is in good agreement
with theoretical estimates of $f \cong 1/3$, as mentioned above.

Properties of stars on the birthline and the relative luminosity of the disc to that of the star,
for $\tilde{f}=0.5$, are summarized in Table~\ref{tabbirthline}, while Fig.~\ref{1348f2}
locates different phases of the star evolution in the HR diagram.

\begin{figure}
   \includegraphics[width=8.8cm]{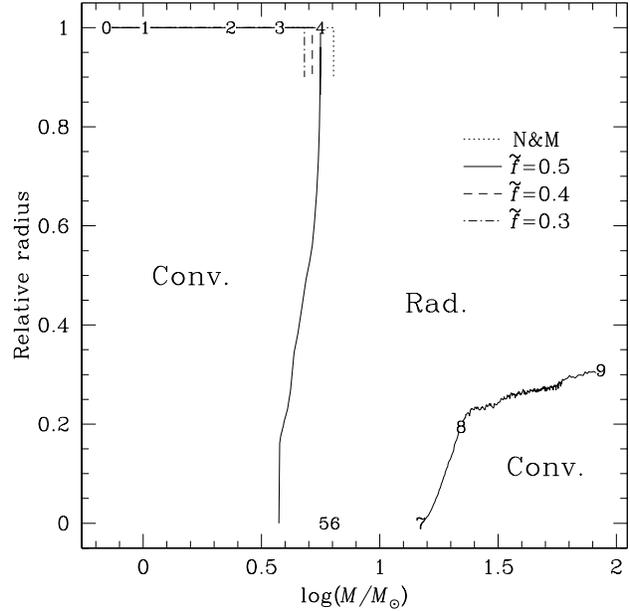}
   \caption{Location of the convective and radiative zones. The relative radius of the
   boundaries between the convective and radiative zones are drawn for the birthline with
   $\tilde{f}=0.5$. For the other birthlines, only the beginning of the fully radiative
   state is shown. The small labels denote the evolutionary stages shown in Fig.~\ref{1348f3}.}
   \label{1348f3}
\end{figure}

\begin{figure}
   \includegraphics[width=8.8cm]{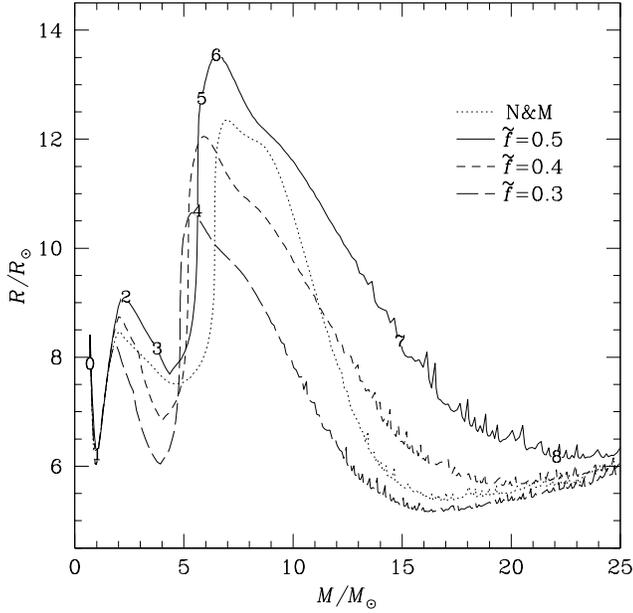}
   \caption{$M$ -- $R$ relation for stars evolving on the birthlines.}
   \label{1348f4}
\end{figure}

The $\ion{^2H}{}$-burning begins at the center of the star. Nascent stars of mass below
$3.8\,M_{\sun}$ ($\log(M/M_{\sun})=0.58$) are fully convective (Fig.~\ref{1348f3}),
and the deuterium accreted is carried inside the
star where it is burnt. As the temperature rises, the deuterium
burns farther from the center. At $3.8\,M_{\sun}$, a radiative core is created,
and the accreted deuterium is then burnt in a shell at the periphery of this core;
the star inflates (second hump in Fig.~\ref{1348f4}) and the luminosity grows.
If the relation of the $L$-parameterized accretion is right, and in particular its single valued
character, the shell $\ion{^2H}{}$-burning could play a major role in the beginning of star formation,
raising the stellar luminosity and inducing via Churchwell's relation, a real boost
in the accretion rate, shortening the time needed to reach high mass stars. The convective $\ion{^2H}{}$-burning
shell becomes thin at the surface of the star around $5.6\,M_{\sun}$
($\log(M/M_{\sun})=0.75$) (slightly depending on the mass accretion rate); the star becomes fully radiative.

The luminosity of the $\ion{^2H}{}$-burning is governed by the accretion rate;
the star inflates to balance a excess of power, or contracts during a shortage of deuterium.
The $R(M)$ relation (Fig.~\ref{1348f4}) is thus very sensitive to $\dot{M}$ during the pre-main
sequence evolution of the star and especially during the fully radiative phase.

\begin{figure}
   \includegraphics[width=8.8cm]{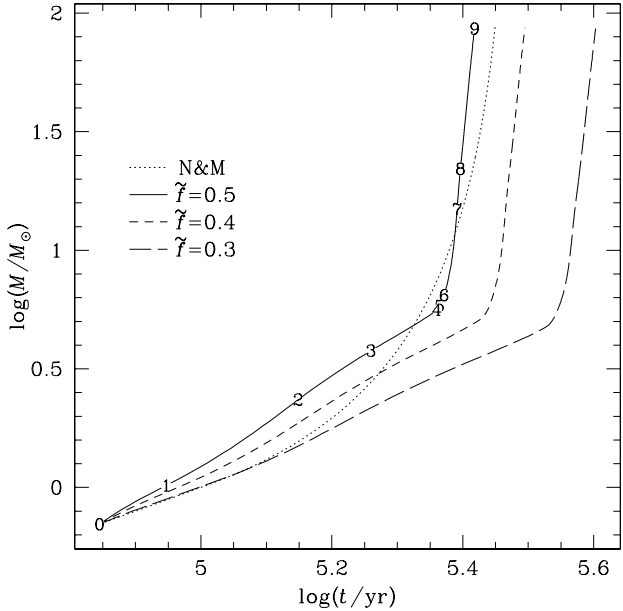}
   \caption{Mass evolution for stars on the birthlines.}
   \label{1348f5}
\end{figure}

The consequences of the luminosity's growth during the shell $\ion{^2H}{}$-burning are clearly visible
in the $\log t$ -- $\log M$ diagram (Fig.~\ref{1348f5}) in the form of an
elbow near $\log(M/M_{\sun})=0.8$ ($6\,M_{\sun}$) with a rather abrupt change of slope in
the curves, for the $L$-parameterized accretion.
Below the bend, the much smoother curve of the $M$-parameterized accretion of N\&M is a little bit higher
than the one for the $L$-parameterized accretion, and lower afterwards. Rather than physical, the origin
of this difference resides in the choice of the parameterization for $\dot{M}$.
Because both laws were adjusted on the same observations, they globally look alike and make similar
predictions, but they can locally differ.
 
\begin{figure}
   \includegraphics[width=8.8cm]{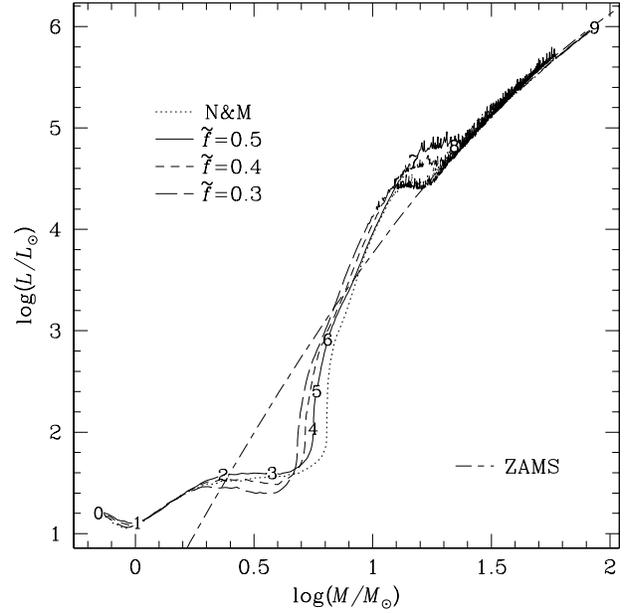}
   \caption{$\log M$ -- $\log L$ relation of stars on the birthlines.}
   \label{1348f6}
\end{figure}

\begin{figure}
   \includegraphics[width=8.8cm]{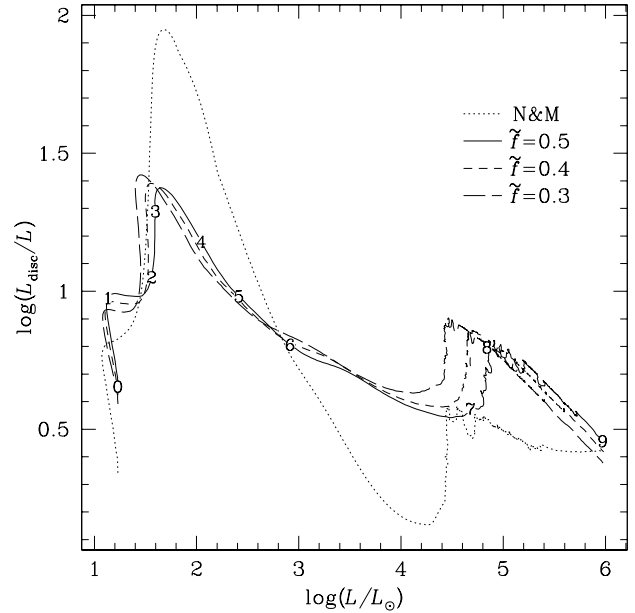}
   \caption{Relative luminosity of disc to that of the star as a function of stellar bolometric luminosity for
   stars evolving on the birthlines. For the model of N\&M, $\tilde{f}=0.5$ is assumed.}
   \label{1348f7}
\end{figure}

One half of the gravitational energy gained during the rotational
infall is transformed into kinetic energy; the other half is used to heat up the disc and is
eventually radiated. Chemical energies (sublimation of dust and grains, ionization) are small
with regards to gravitational
energy and are neglected in the computations. The disc luminosity,
$L_\mathrm{disc}$, is thus of the order of $G\,M\,\dot{M}_{disc}\,/\,(2\,R)$, as given by
Hartmann (\cite{hart1998a}, \S 5.4).
It appears in Fig.~\ref{1348f7} that during the accretion phase,
the luminosity of the disc is always greater than the star luminosity.
On the birthline, the star encounters three phases of nearly constant luminosity. In these
phases, the radius of the star shrinks while the mass continues to grow; according to the
previous formula, the luminosity of the disc grows. That explains the three vertical steps
in the $\log L$ -- $\log(L/L_\mathrm{disc})$ diagram. The step
around $\log(L/L_{\sun})=4.8$ corresponds to the arrival of the star on the ZAMS and is
the only feature really sensitive to $\tilde{f}$. The probability of having good observations of
stars in this region of the diagram is very small and we conclude that the relative
luminosity of the disc to that of the star cannot help us to fix $\tilde{f}$.
The smaller variability of the ratio of disc to star luminosities for the
$L$-parameterized accretion than for the $M$-parameterized accretion has the same origin as
in the $\log t$ -- $\log M$ diagram.

After the accreting star has joined the ZAMS, at $22\,M_{\sun}$, it evolves in a very similar manner to the one
calculated in N\&M. They used $\dot{M}\propto M^{1.5}$ and the linear fit in Churchwell's
data implies $\dot{M}\propto M^{1.54}$ using the $L(M)$ relation on the ZAMS.
An accretion rate of $8 \cdot 10^{-3}\,M_{\sun}\,\mathrm{yr}^{-1}$ is obtained for a $85\,M_{\sun}$ star.
Though it seems very high, this rate is still in the permitted domain of accretion with
spherical symmetry (Wolfire \& Cassinelli \cite{wolcas}); if the accretion is made via a disc,
even higher rates are permitted due to the less important cross-section of infalling material
to radiations from the star and from the shock, and to the enhanced cooling
(Nakano \cite{nakaal3}; Henning \cite{henning}).

Once the disc is no longer fed by its environment, for example by the disappearance of the parent cloud by
the outflows of the star or jets/ouflows and the radiation of neighbouring hot luminous stars (Nakano et al. \cite{nakaal1}),
the disc get thinner by accretion and evaporation. The transition from active to passive discs
is probably a smooth function of time. As its mass decreases, the residual disc plays a
smaller and smaller role in the evolution of the star.
Both the mass of discs and the typical duration of the transition are still very uncertain;
in the computations, we neglect their effects, using an instantaneous stopping of the
accretion at time $t_{\mathrm{stop}}$.
Figure~\ref{1348f8} shows the resulting theoretical tracks for some values of
$t_\mathrm{stop}$ (in units of $10^3\,\mathrm{yr}$).

The following considerations can partly justify the assumption that from a certain stage,
the disc has no more significant influence on the central star.
For low and intermediate mass stars, the disc is usually lighter
than the central star (Hartmann \cite{hart1998a}, \S\S 6.3 and 9.7, \cite{hart1998b}; Bernasconi \cite{bernasconi}).
Observations of the residual mass transfer rate of the disc around T Tauri stars is
described by Hartmann (\cite{hart1998b}); they can be approximately represented by
$\log \dot{M}_{\mathrm{disc}}^{\mathrm{res}} \approx
0.4-1.4\,\log \tilde{t}$ (units are $M_{\sun}\,\mathrm{yr}^{-1}$ and $\mathrm{yr}$), where
$\tilde{t}$ is a relative scale for the ages.
Integrating $\dot{M}_{\mathrm{disc}}^{\mathrm{res}}$ from
the $\tilde{t}$ corresponding to a disc mass transfer rate
$(1+\tilde{f})\,\dot{M}_{\mathrm{out}}$ at $t_{\mathrm{stop}}$ to
infinity, one finds a crude estimate of the mass of the disc at the moment of the
stopping of the accretion:
$M_{\mathrm{disc}} \approx 5\,\dot{M}_{\mathrm{disc}}{}^{0.3}$
(units are $M_{\sun}$ and $M_{\sun}\,\mathrm{yr}^{-1}$). $\dot{M}_{\mathrm{disc}}$ is
probably out of the validity domain for $\dot{M}_{\mathrm{disc}}^{\mathrm{res}}$,
but $f\,M_{\mathrm{disc}}$ remains less than $5\%$ of the mass of the star,
for $M > 2\,M_{\sun}$. Bernasconi (\cite{bernasconi}) studied the residual mass accretion of dying discs,
but his models contain many arbitrary parameters; for intermediate mass stars, the difference between
the tracks with and without residual accretion are nonetheless much smaller than the error bars of the
present day observations.

The time $t_{\mathrm{stop}}$ corresponds approximately to the time at which the cocoon
of interstellar matter vanishes, thus the star becomes visible in the optical bands only after this
moment; the location of the first optical visibility of stars above $22\,M_{\sun}$ is very close to the ZAMS.

For the stars studied in this paper (mass on the ZAMS $>1.19\,M_{\sun}$),
there are four regions in the pre-main sequence part of the HR diagram, as shown in
Fig.~\ref{1348f8}. 1) For $\log T_{\mathrm{eff}}$ below $\approx 3.68$ (i.e. $4.8\cdot 10^3\,\mathrm{K}$),
the stars are fully convective; they leave the birthline with a vertical Hayashi-like
displacement in the HR
diagram. No more accreting stars between $\log T_{\mathrm{eff}} \approx 3.68$ and $\log T_{\mathrm{eff}}
\approx 3.76$ (i.e. $5.8 \cdot 10^3\,\mathrm{K}$) have a radiative core surrounded by a
convective $\ion{^2H}{}$-burning shell; if the star already has a radiative core before
the end of accretion, the track closely follows the birthline (with an increasing luminosity) for a short period of time after
$t_\mathrm{stop}$; if a radiative core is absent, one is swiftly created after the
end of accretion and the star leaves the birthline with a decreasing luminosity.
3) For $\log T_{\mathrm{eff}}$ above $\approx 3.76$, the stars are fully radiative, except
on 4) a small band above the ZAMS where stars already have a convective $\ion{H}{}$-burning core.

Electronic tables for the tracks are available via the anonymous ftp server of the Geneva
observatory; the file is located in directory\newline
$ftp://obsftp.unige.ch/pub/evol/prems/AA2001BM$ .


\section{Relation with the IMF}

The initial mass function (IMF) represents the density
of probability that a star ends its accretion with a given mass.
The IMF results from many physical phenomena (fragmentation of the parent cloud, vanishing of the
reservoir of interstellar matter, accretion rate and its efficiency, binarity, etc.).
But under the assumption that the accretion rate does not depend on the local conditions
of the cloud and disc, the IMF can be seen as the result of the mechanisms that end the
feeding of the disc with interstellar matter (in the case of a small ratio of disc to star
masses).

\begin{figure}
   \includegraphics[width=8.8cm]{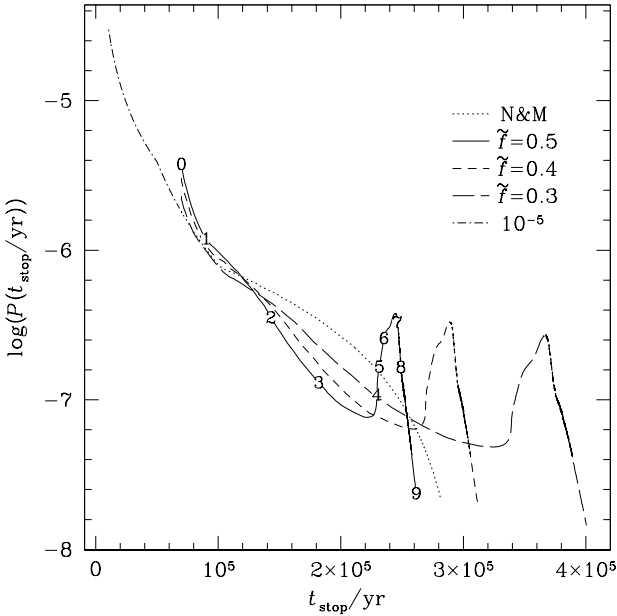}
   \caption{Density of probability of $t_\mathrm{stop}$ for mass accretion rates
   with three values of $\tilde{f}$, and the N\&M one. For $t_{\mathrm{stop}}<7 \cdot 10^4\,\mathrm{yr}$ ($M<0.7\,M_{\sun}$), an accretion rate of
   $10^{-5}\,M_{\sun}\,\mathrm{yr}^{-1}$ is assumed.}
   \label{1348f9}
\end{figure}

The IMF is usually noted $\xi(m)$. $\xi(m)\,\mathrm{d}m$ is the relative number
of new-born stars in the mass interval $\mathrm{d}m$ centered on $m$; the IMF is
normalized so that $\int_{M_{\min}}^{M_{\max}}\,\xi(m)\,\mathrm{d}m=1$.
The assumptions made for the models of this paper imply that the star mass evolution
on the birthline is a single valued and monotonously increasing function $M(t)$.
Thus, from the observed IMF and using the $M(t)$ relation (as in Fig.~\ref{1348f4}),
one can compute the density of probability $P(t_\mathrm{stop})$ of the age $t_\mathrm{stop}$ of stars
when the accretion is abruptly stopped:
\[P(t_\mathrm{stop})=(\xi(M(t))\,\mathrm{d}M(t)/\mathrm{d}t)|_{t=t_\mathrm{stop}}.\]
The result is shown in Fig.~\ref{1348f9}, using the IMF of Kroupa (\cite{kroupa}):
$\xi(m)=\beta\,m^{-\alpha}$ where \[\left(\alpha;\,\beta \right)=\left\{
\begin{array}{cl}
\left(0.30;\,1.99 \right), & $if $m \in [M_{\min}=0.01,\,0.08), \\ 
\left(1.30;\,0.159 \right), & $if $m \in [0.08,\,0.50), \\ 
\left(2.30;\,0.0795 \right), & $if $m \in [0.50,\,M_{\max}=100.]
\end{array}\right.\]
The IMF of Scalo (\cite{scalo}) differs from Kroupa's one principally in the
$<0.1\,M_{\sun}$ region; both IMFs give similar results.

For the $L$-parameterized accretion, the break in the mass evolution graph located
near $6\,M_{\sun}$ is great enough to generate a sharp hump (an order of magnitude) in
the density of probability of $t_\mathrm{stop}$,
$P(t_\mathrm{stop})$. For the $M$-parameterized accretion of N\&M which gives a monotonously
decreasing $P(t_\mathrm{stop})$, a bump is also
distinguishable, but it is smoothed by the nature of the law used. In
comparison to the $L$-parameterized accretion rate, the $M$-parameterized accretion rate
is indeed
a little higher for masses in the range $1.6$ to $6\,M_{\sun}$ and lower for the other stars
(Fig.~\ref{1348f5}).
The presence of the peak for the $L$-parameterized accretion does not depend on the ratio
$\tilde{f}=\dot{M}/{\dot{M}_{\mathrm{out}}}$ of accreted and ejected masses, if this quantity
does not evolve with the mass and luminosity of the star.
If theses peaks are confirmed, it would be most interesting to investigate the various
physical processes at their origin.
One possible way to explain theses curves could be as follows. For stars below $6\,M_{\sun}$, the
fractal parent cloud is probably dissipated mostly by neighbouring stars. For stars above
$6\,M_{\sun}$, the
star and disc luminosities could grow high enough to scatter the parent cloud.

\section{Conclusions}

   Numerical experiments have been done with mass accretion rates proportional to the
   luminosity parameterized outflow rate
   \[\log{{\dot{M}_\mathrm{out}} \over {M_{\sun}\,\mathrm{yr}^{-1}}}=-5.28+\log {L
   \over{L_{\sun}}} \cdot(0.752-0.0278\log {L \over{L_{\sun}}})\]
   established by Churchwell
   (\cite{church}) and confirmed by Henning et al. (\cite{henning2000}).
   Until complete theoretical values of the ratio of
   accreted on the star and ejected masses, $\tilde{f}=\dot{M}/\dot{M}_{\mathrm{out}}$,
   are computed taking into account stellar rotation
   and magnetic field, chemistry, radiative transfers, etc., $\tilde{f}$ can be seen as a
   unknown parameter.
   \begin{enumerate}
      \item The comparison of theoretical birthlines from different $\tilde{f}$ to
      observations of young star constraints $\tilde{f}$. For these experiments, we chose
      constant values for $\tilde{f}$ and our favorite choice is $\tilde{f}=0.5$. This value permits
      a reasonable fit of the observations of accreting stars.
      \item Only a fraction $f=\tilde{f}/(1+\tilde{f})=1/3$ of the mass transferred
      from the disc in the central region accretes on the star. This ratio is in
      relative agreement with numerical predictions by Tomisaka
      (\cite{tomisaka1}) ($1/3$) and the theoretical model of Shu et al. (\cite{shuetal})
      (also $1/3$), both cases are for low mass stars. Observations by Churchwell (\cite{church})
      suggest a smaller value (of the order of $0.15$) for a B star; this is not seen as a strong disagreement.
      \item The relative lack of observations in the upper pre-main sequence part of the HR diagram
      is due to the very fast evolution of the stars just before $\ion{^1H}{}$-ignition
      and their arrival on the ZAMS; there are few chances to better constrain $\tilde{f}$
      using the HR diagram. Apparently, the ratio of disc to star luminosities cannot help
      to fix $\tilde{f}$.
      \item The $R(M)$ relation is very sensitive to the mass accretion rate during the
      $\ion{^2H}{}$-burning; the biggest differences are located in the range of $7$ to
      $15\,M_{\sun}$, i.e. when the star is fully radiative and the $\ion{^2H}{}$ burns
      near the surface; for a given mass, the greater the rate, the bigger
      the radius.
      \item The $M$-parameterized accretion rate of Norberg \& Maeder (\cite{normae}) gives
      essentially the same birthline in the HR diagram as the $L$-parameterized accretion.
      The mass evolution of stars with accretion rate parameterized by $M$ is smooth in $\log t$ -- $\log M$;
      on the contrary, the evolution for the mass with the $L$-parameterized accretion
      shows a sharp transition near $6\,M_{\sun}$.
      This is related to the increasing radius and luminosity of the star during the
      shell $\ion{^2H}{}$-burning.
      This dramatic change, as seen in the diagram of Fig.~\ref{1348f9},
      has certainly a counterpart in the shape of the IMF. Kroupa (\cite{kroupa}) guesses
      that the slope of the observed IMF could have a break near $6\,M_{\sun}$, just in
      the elbow of $M(t)$.
      \item At $85\,M_{\sun}$, the accretion rate is about $8 \cdot
      10^{-3}\,M_{\sun}\,\mathrm{yr}^{-1}$. This tremendous rate is not forbidden by star and shock luminosities,
      even in the unfavorable spherical case studied by Wolfire \& Cassinelli (\cite{wolcas}).
      \item The time needed to form a $85\,M_{\sun}$ star is about one quarter of $\mathrm{Myr}$ for $\tilde{f}=0.5$ and
      one third of $\mathrm{Myr}$ for $\tilde{f}=0.3$. In all cases, it is much shorter than the hydrogen burning timescale.
   \end{enumerate}

   Growing accretion seems to be a valuable scenario to explain the formation of massive
   stars, but the field for theories about the physical backgrounds
   of the accretion processes is still widely open. There are open questions for which
   this prospective work could not give answers. Is the hypothesis of the constancy of $\tilde{f}$ valid
   for all stars on the birthline ? How can one explain such a high $\dot{M}$ ?
   Is $\dot{M}(L)$ really single valued, as suggested by the observations ? 
   What occurs in the accreting stars observed far from the birthline (for $5<\log(L/L_{\sun})<6$) ? 
   Observational, theoretical and numerical works should
   be done in order to better understand the mechanisms of the huge accretion processes.

\begin{acknowledgements}
The authors express their thanks to Dr.~G.~Meynet and to P.~Norberg for their help
and fruitful discussions during the use of the Geneva stellar evolution code.

\end{acknowledgements}

\end{document}